\documentclass[aps,prl,preprint,tightenlines,superscriptaddress,showpacs,byrevtex]{revtex4}
\usepackage{epsfig}

\newcommand{\mpl}{M_{p\bar{\Lambda}}}
\newcommand{\mb}{{M_{\rm bc}}}
\newcommand{\de}{{\Delta{E}}}
\newcommand{\plpi}{{p\bar{\Lambda}\pi^-}}
\newcommand{\plk}{{p\bar{\Lambda}K^-}}
\newcommand{\psigpi}{{p\bar{\Sigma}^0\pi^-}}

\begin{document}




\epsfysize 25mm
\epsfbox{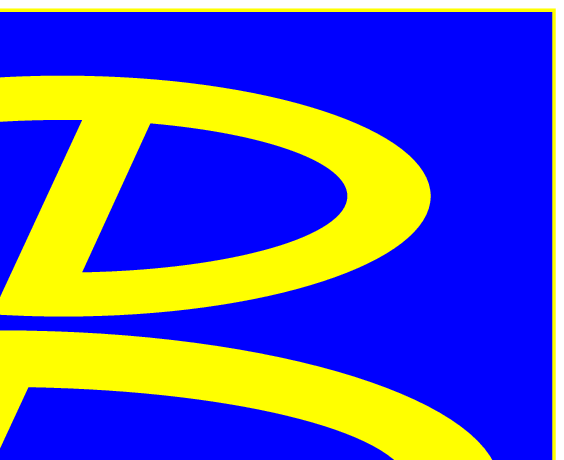}
\begin{flushright}
\vskip -25mm
\noindent
\hspace*{3.0in}{\bf Belle preprint 2003-2} \
\end{flushright}
\vskip 25mm

\title{ \quad\\[1cm] \Large
Observation of $B^0 \to \plpi$}

\affiliation{Budker Institute of Nuclear Physics, Novosibirsk}
\affiliation{Chuo University, Tokyo}
\affiliation{University of Cincinnati, Cincinnati, Ohio 45221}
\affiliation{University of Hawaii, Honolulu, Hawaii 96822}
\affiliation{High Energy Accelerator Research Organization (KEK), Tsukuba}
\affiliation{Hiroshima Institute of Technology, Hiroshima}
\affiliation{Institute of High Energy Physics, Vienna}
\affiliation{Institute for Theoretical and Experimental Physics, Moscow}
\affiliation{J. Stefan Institute, Ljubljana}
\affiliation{Kanagawa University, Yokohama}
\affiliation{Korea University, Seoul}
\affiliation{Kyoto University, Kyoto}
\affiliation{Kyungpook National University, Taegu}
\affiliation{Institut de Physique des Hautes \'Energies, Universit\'e de Lausanne, Lausanne}
\affiliation{University of Ljubljana, Ljubljana}
\affiliation{University of Maribor, Maribor}
\affiliation{University of Melbourne, Victoria}
\affiliation{Nagoya University, Nagoya}
\affiliation{National Kaohsiung Normal University, Kaohsiung}
\affiliation{National Lien-Ho Institute of Technology, Miao Li}
\affiliation{Department of Physics, National Taiwan University, Taipei}
\affiliation{H. Niewodniczanski Institute of Nuclear Physics, Krakow}
\affiliation{Nihon Dental College, Niigata}
\affiliation{Niigata University, Niigata}
\affiliation{Osaka City University, Osaka}
\affiliation{Osaka University, Osaka}
\affiliation{Panjab University, Chandigarh}
\affiliation{Peking University, Beijing}
\affiliation{Saga University, Saga}
\affiliation{University of Science and Technology of China, Hefei}
\affiliation{Seoul National University, Seoul}
\affiliation{Sungkyunkwan University, Suwon}
\affiliation{University of Sydney, Sydney NSW}
\affiliation{Tata Institute of Fundamental Research, Bombay}
\affiliation{Toho University, Funabashi}
\affiliation{Tohoku Gakuin University, Tagajo}
\affiliation{Tohoku University, Sendai}
\affiliation{University of Tokyo, Tokyo}
\affiliation{Tokyo Institute of Technology, Tokyo}
\affiliation{Tokyo Metropolitan University, Tokyo}
\affiliation{Tokyo University of Agriculture and Technology, Tokyo}
\affiliation{Toyama National College of Maritime Technology, Toyama}
\affiliation{University of Tsukuba, Tsukuba}
\affiliation{Utkal University, Bhubaneswer}
\affiliation{Virginia Polytechnic Institute and State University, Blacksburg, Virginia 24061}
\affiliation{Yonsei University, Seoul}
\author{M.-Z.~Wang}\affiliation{Department of Physics, National Taiwan University, Taipei} 
  \author{Y.-J.~Lee}\affiliation{Department of Physics, National Taiwan University, Taipei} 
  \author{K.~Abe}\affiliation{High Energy Accelerator Research Organization (KEK), Tsukuba} 
  \author{K.~Abe}\affiliation{Tohoku Gakuin University, Tagajo} 
  \author{T.~Abe}\affiliation{Tohoku University, Sendai} 
  \author{H.~Aihara}\affiliation{University of Tokyo, Tokyo} 
  \author{M.~Akatsu}\affiliation{Nagoya University, Nagoya} 
  \author{Y.~Asano}\affiliation{University of Tsukuba, Tsukuba} 
  \author{T.~Aso}\affiliation{Toyama National College of Maritime Technology, Toyama} 
  \author{T.~Aushev}\affiliation{Institute for Theoretical and Experimental Physics, Moscow} 
% ...
\collaboration{The Belle Collaboration}

\tighten




\normalsize

\begin{abstract}

We report the first observation of 
the charmless hyperonic $B$ decay, $B^0 \to \plpi$, 
using a 78 fb$^{-1}$ data sample recorded 
on the $\Upsilon({\rm 4S})$ resonance with 
the Belle detector at KEKB.
The measured branching fraction is
${\mathcal B}(B^0 \to \plpi) = (3.97 ^{+1.00}_{-0.80} \pm 0.56) 
\times 10^{-6}$. 
Searches for $B^0 \to \plk$ and $\psigpi$ yield 
no significant signals 
and we set 90\% confidence-level upper limits of
${\mathcal B}(B^0 \to \plk) < 8.2 \times 10^{-7}$ and
${\mathcal B}(B^0 \to \psigpi) < 3.8 \times 10^{-6}$. 

\vskip1pc

\end{abstract}
%
%
%

\maketitle

{\renewcommand{\thefootnote}{\fnsymbol{footnote}}

\setcounter{footnote}{0}

\normalsize

%
%


The Belle collaboration recently reported 
the observation of $B^+ \to p\bar{p}K^+$~\cite{ppk}, 
which is the first known example of $B$ meson
decay to charmless final states containing baryons. 
The three-body decay rate is larger than the rate for two-body decays
(such as $B\to p\bar p$~\cite{2body}),
and the observed $M_{p\bar p}$ spectrum peaks near threshold, 
in agreement with theoretical suggestions~\cite{HS,rhopn}.
In this Letter we report the first observation of 
the related three-body decay $B^0\to \plpi$, 
and a search for $B^0\to \plk$ and $\psigpi$ modes.
The rate for $B^0\to \plpi$ is comparable to
$B^+ \to p\bar{p}K^+$, and 
the observed $M_{p\bar \Lambda}$ spectrum 
again peaks toward threshold.

In the Standard Model, 
these decays proceed via $b\to u$ tree and $b\to s (d)$ penguin diagrams.  
They may be used to 
search for direct $CP$ violation and 
test our theoretical understanding of 
rare decay processes involving baryons~\cite{HS,rhopn,PP,CY,CHT}.
Modes involving hyperons, in particular,
can probe the $s$ quark chirality in $B$ decay~\cite{suzuki};
with sufficient statistics,
they could provide a tool for probing $T$ violation~\cite{HS}
via the self-analyzed $\Lambda$ polarization information.

We use a  78 fb$^{-1}$  data sample,
consisting of $85.0 \pm 0.5$ million $B\overline{B}$ pairs,
collected by the Belle detector 
at the KEKB asymmetric energy $e^+e^-$ (3.5 on 8~GeV) collider. 
The Belle detector is a large-solid-angle magnetic spectrometer 
that consists of a three-layer silicon vertex detector (SVD),
a 50-layer central drift chamber (CDC), an array of
aerogel threshold \v{C}erenkov counters (ACC), 
a barrel-like arrangement of time-of-flight
scintillation counters (TOF), and an electromagnetic calorimeter
comprised of CsI(Tl) crystals located inside 
a super-conducting solenoid coil that provides a 1.5~T
magnetic field.  An iron flux-return located outside of
the coil is instrumented to detect $K_L^0$ mesons and to identify
muons.  The detector
is described in detail elsewhere~\cite{Belle}.

Since the $e^+e^-$ center-of-mass energy is set to match 
the $\Upsilon({\rm 4S})$ resonance, 
which decays into a $B\overline{B}$ pair, one
can use the following two kinematic variables 
to identify the reconstructed $B$ meson candidates~\cite{conjugate}: 
the beam-energy constrained mass, $\mb =
\sqrt{E^2_{\rm beam}-p^2_B}$, and the energy difference, $\de =
E_B - E_{\rm beam}$, where $E_{\rm beam}$ is the beam energy, and
$p_B$ and $E_B$ are the 
momentum and energy of the reconstructed $B$ meson in the  
$\Upsilon({\rm 4S})$ rest frame. 
The candidate region is defined as
5.2 GeV/$c^2 < \mb < 5.29$ GeV/$c^2$ 
and $|\de|< 0.2$ GeV in this analysis.

The event selection criteria are based on the information obtained
from the tracking system 
(SVD+CDC) and the hadron identification system (CDC+ACC+TOF), 
and are optimized using Monte Carlo (MC)
simulated event samples.

All primary charged tracks 
are required to satisfy track quality criteria
based on the track impact parameters relative to the   
interaction point (IP). 
The deviations from the IP position are required to be within
$\pm$0.3 cm in the transverse ($x$-$y$) plane, and within $\pm$3 cm
in the $z$ direction, where the $z$ axis is 
defined by the positron beam line.
Primary proton candidates are selected based
on $p/K/\pi$ likelihood functions
obtained from the hadron identification system. 
We require
$L_p/(L_p+L_K)> 0.3 $ and  $L_p/(L_p+L_{\pi})> 0.6$,
where $L_{p/K/\pi}$ stands for the proton/kaon/pion likelihood.
For kaons (pions), we
require the kaon (pion) $K$-$\pi$ likelihood ratio
to be greater than 0.6.  
$\Lambda$ candidates are reconstructed via the $p\pi^-$ decay channel 
using the method described in Ref.~\cite{2body}. 
$\Sigma^0$ candidates are reconstructed via the $\Lambda\gamma$ decay channel,
where we use a 35 MeV/$c^2$ mass window around the nominal mass~\cite{PDG} and 
require the  $\gamma$ energy to be greater than 100 MeV.

The dominant background for the rare decay modes 
reported here 
is from the continuum $e^+e^- \to q\bar{q}$ process.
The background from $B$ decays is much smaller.
This is confirmed with
an off-resonance data set (8.8 fb$^{-1}$) 
accumulated at an energy that is 60 MeV 
below the $\Upsilon({\rm 4S})$, and an
MC sample of 120 million continuum events.
In the $\Upsilon({\rm 4S})$ rest frame, 
continuum events are jet-like  while
$B\overline{B}$ events are spherical. 
We follow the scheme defined in Ref.~\cite{etapk} and  
combine seven shape variables to form 
a Fisher discriminant~\cite{fisher} that is used to optimize   
continuum background suppression. The variables chosen have 
almost no correlation
with $\mb$ and $\de$.
Probability density functions (PDF's) for the Fisher discriminant and 
the cosine of the angle between the $B$ flight direction 
and the $e^-$ beam direction in the $\Upsilon({\rm 4S})$ rest frame
are combined to form the signal (background)
likelihood ${\cal L}_{\rm S (BG)}$.
We require   
the likelihood ratio ${\cal LR} = {\cal L}_{\rm S}/({\cal L}_{\rm
S}+{\cal L}_{\rm BG})$ to be greater than 0.8,  
which suppresses about 94\% of the background while retaining 
66\% of the signal.
The signal and background PDF's are obtained from MC simulation studies.


\begin{figure}[htb]
\centerline{
\epsfig{file=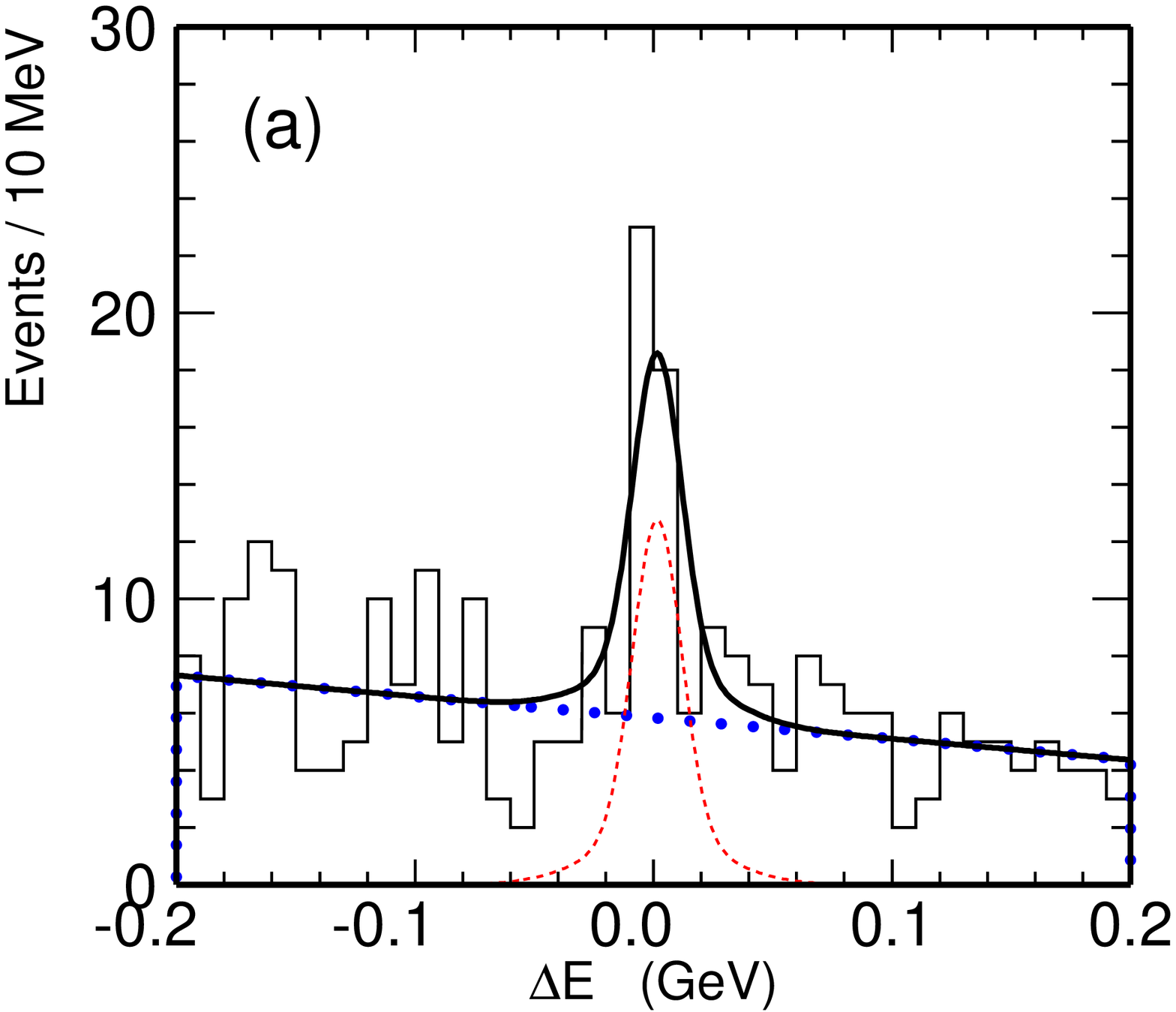,width=2.55in}
\epsfig{file=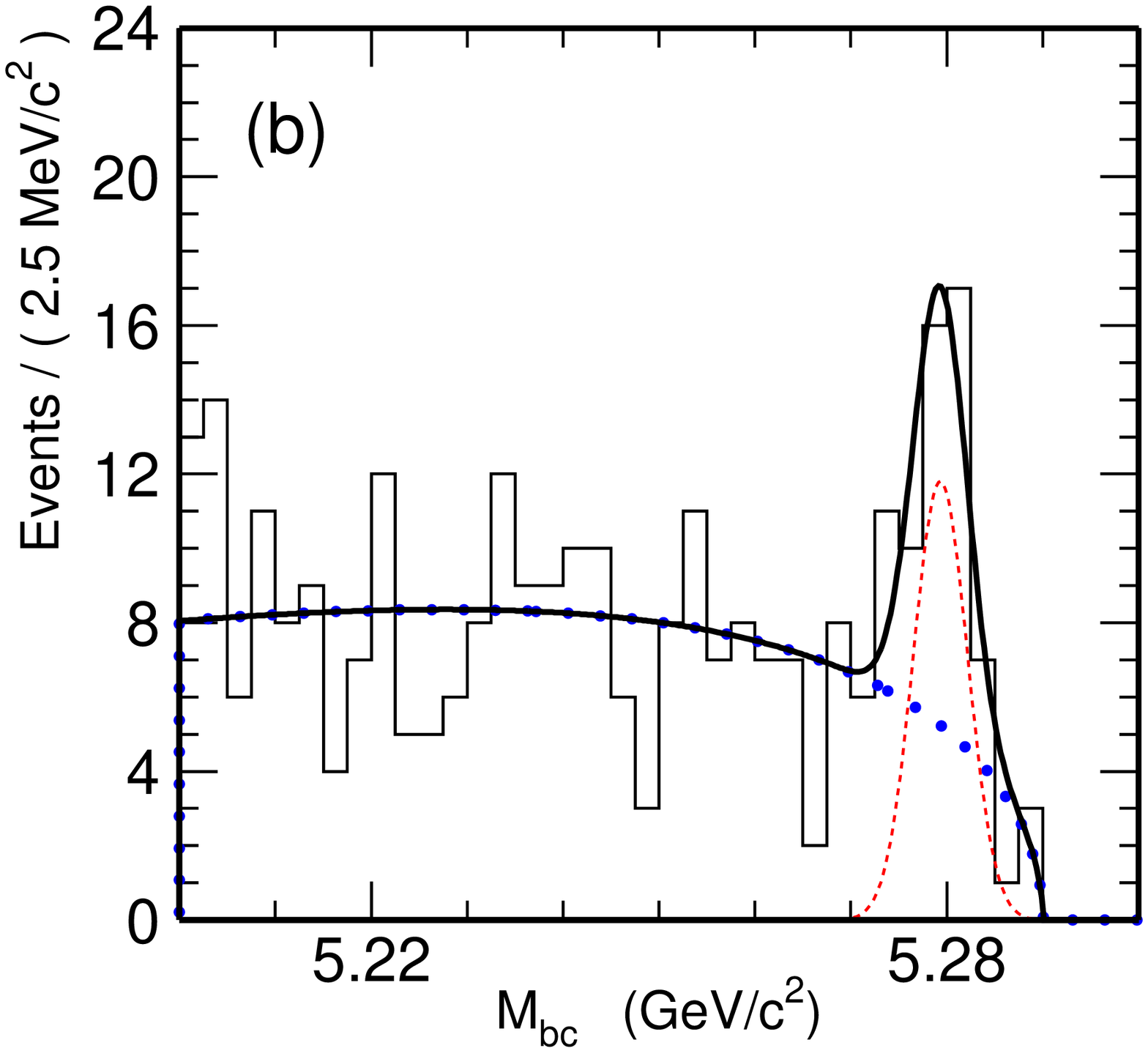,width=2.482in}    
}
\caption{The (a) $\de$ and (b) $\mb$
distributions for $B^0 \to \plpi$ candidates. 
The solid, dotted and dashed lines represent
the combined fit result, fitted background and fitted signal, respectively.}   
\label{fg:plpimbde}
\end{figure}

Figure~1(a) shows the $\de$ distribution for
selected $\plpi$ candidates that have
$\mb >$ 5.27 GeV/$c^2$;
Fig.~\ref{fg:plpimbde}(b) shows the  $\mb$ distribution
for events with $|\de| <$ 0.03 GeV.
With the current statistics, no intermediate
resonances are evident in the Dalitz plot for
this channel.
We use a binned likelihood fit to estimate the signal yield. 
A Gaussian is used to parameterize the signal
in $\mb$ while a double Gaussian is used for $\de$. The 
Gaussian parameters are determined from MC simulation.
Background shapes are studied using side-band
events (0.1 GeV $ < |\de| < 0.2$ GeV for the $\mb$ study and 
5.20 GeV/$c^2$ $ < \mb <$ 5.26 GeV/$c^2$ for $\de$  
) and are checked with the continuum MC sample.
We use the ARGUS function~\cite{Argus} to model
the $\mb$ background 
 and a linear function for the
$\de$ background.
The fit results are shown as curves in Fig.~\ref{fg:plpimbde}.
The fit to the $\de$ distribution yields
39.2 $^{+9.1}_{-8.4}$ candidates with a significance of $5.8$ standard 
deviations.
The fit to the $\mb$ distribution yields
$33.7^{+8.1}_{-7.4}$ candidates with a significance of $5.7$ standard 
deviations. 
The smaller yield in the $\mb$ fit is consistent with the $\de$
fit result after taking into account the efficiency of the
$|\de|< 0.03$ GeV selection.
The signal yields and the branching fractions
are determined from fits to the $\Delta E$ distribution
rather than to $\mb$ in order to minimize
possible bias from $B\bar{B}$ background, which tends to
peak in $\mb$ but not in $\de$.

Since the decay is not uniform in phase space,
we fit the $\Delta E$ signal yield in
bins of $M_{p \bar{\Lambda}}$,
and correct for the MC-determined detection
efficiency for each bin.
This reduces the  model dependence of the branching
fraction determination. 
The signal yield as a function
of $p\bar{\Lambda}$ mass is shown in Fig.~\ref{fg:phase}. 
The distribution from a three-body phase space
MC, normalized to the area of
the signal, is superimposed. 
The observed mass distribution peaks at low $p\bar{\Lambda}$ mass,
similar to that observed for  $B^+\to p\bar p K^+$ decays~\cite{ppk}.
The results of the fits, along with the efficiencies and 
partial branching fractions for each $\mpl$ bin, are given in Table~\ref{bins}.
We sum the partial branching
fractions in Table~\ref{bins} to obtain
\[
{\cal B}(B^0 \to \plpi) =(3.97\,^{+\,1.00}_{-\,0.80} \; ({\rm stat})
           \pm 0.56 \; ({\rm syst})) \times 10^{-6},
\]
where the systematic uncertainty is described below.

\begin{table}[htb]
\caption{The event yield, efficiency, and branching
fraction (${\cal B}$)              
for each  $M_{p\bar{\Lambda}}$ bin.}
\label{bins}
\begin{center}
\begin{tabular}{cccc}
$M_{p\bar{\Lambda}}$ ( GeV/$c^2$)&
signal yield&
efficiency(\%)&
${\cal B}$ ($10^{-6}$)
\\
\hline
$<2.2$&
$11.4^{+4.0}_{-3.3}$&
12.5&
$1.08^{+0.37}_{-0.31}$
\\
\hline
$2.2-2.4$&
$11.2^{+4.4}_{-3.7}$&
11.7&
$1.12^{+0.44}_{-0.37}$
\\
\hline
$2.4-2.6$&
$2.4^{+2.7}_{-2.0}$&
11.1&
$0.25^{+0.29}_{-0.21}$
\\
\hline
$2.6-2.8$&
$2.4^{+2.6}_{-1.8}$&
9.9&
$0.28^{+0.31}_{-0.22}$
\\
\hline
$2.8-3.4$&
$2.4^{+2.9}_{-2.2}$&
11.4&
$0.24^{+0.30}_{-0.23}$
\\
\hline
$3.4-4.0$&
$5.0^{+3.6}_{-2.8}$&
11.7&
$0.51^{+0.36}_{-0.29}$
\\
\hline
$4.0-4.6$&
$-3.3^{+2.3}_{-1.8}$&
12.5&
$-0.31^{+0.21}_{-0.17}$
\\
\hline
$>4.6$&
$7.0^{+4.2}_{-3.5}$&
10.4&
$0.79^{+0.48}_{-0.39}$
\\
\end{tabular}
\end{center}
\end{table}

\begin{figure}[htb]
\centering
\mbox{\epsfig{figure=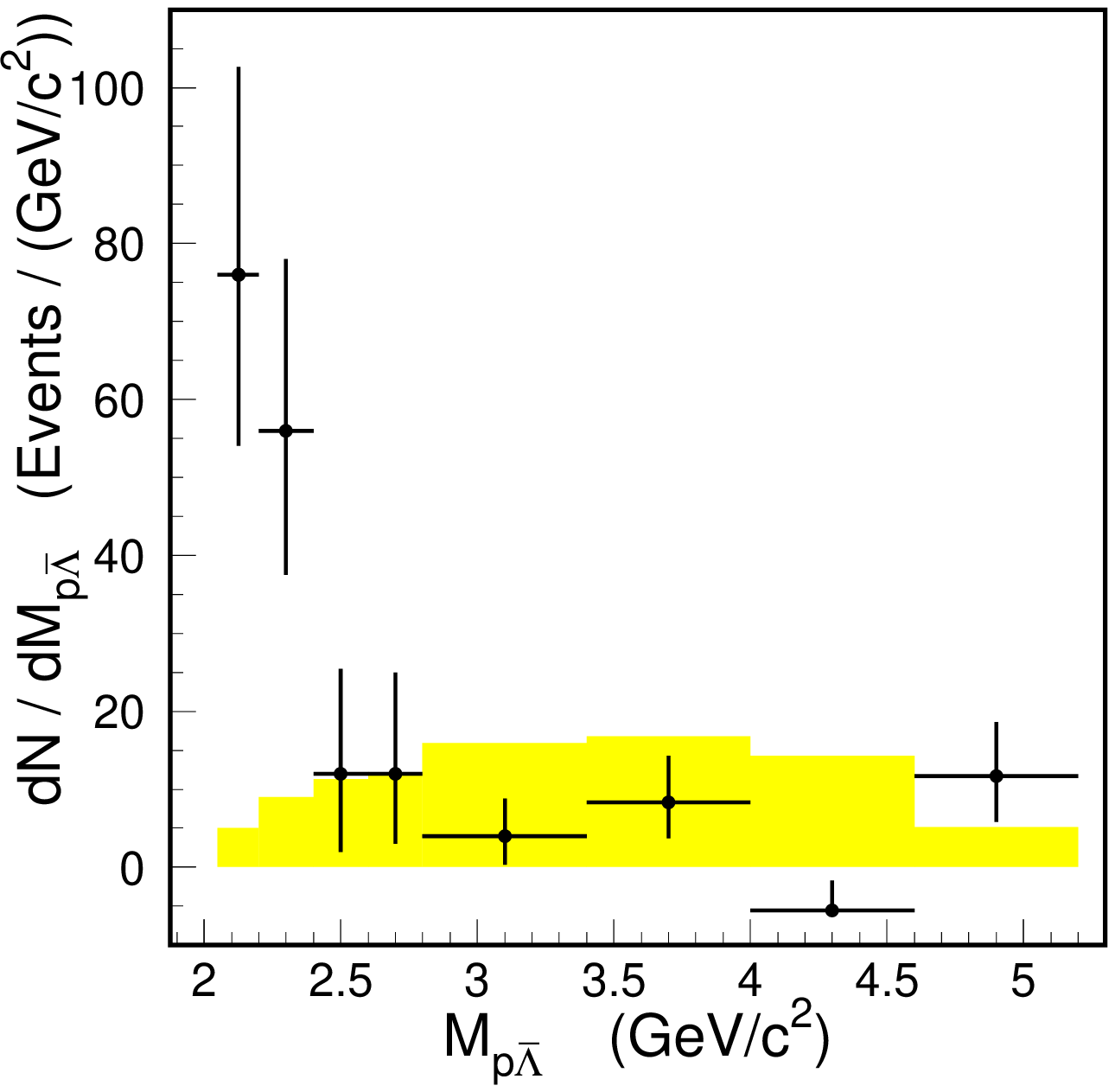,width=3.in}}
\centering
\caption{The fitted yield divided by the bin size for
$B^0\to \plpi$ as a function of $M_{p \bar{\Lambda}}$.
The shaded distribution is from a phase-space MC simulation 
with area normalized to
signal yield.}
\label{fg:phase}
\end{figure}

The systematic uncertainty in particle selection  
is studied mainly using high statistics control samples.
Proton identification
is studied with a  $\Lambda \to p \pi^-$ sample. 
Kaon/pion identification is studied with a $D^{*+} \to D^0\pi^+$,
 $D^0 \to K^-\pi^+$ sample. 
Tracking efficiency is studied with 
$\eta \to \gamma\gamma$ and $\eta \to \pi^+\pi^-\pi^0$ samples.
Based on these studies, 
we assign a 2\% error for each track, 3\% for each proton identification
requirement,
and 2\% for each kaon/pion identification requirement.

We study the $\cal LR$ continuum suppression
by varying the $\cal LR$ cut value from 0 to 0.9 to check the systematic
trend. The systematic error is found to be 4\%.
The additional uncertainty of off-IP tracks  
for $\Lambda$ reconstruction is estimated to be 6\%,
which is determined from the difference
of the proper decay time distributions for data and MC simulation. 

The systematic uncertainty  in the fit yield is
 studied by varying the parameters of
the signal and background PDF's. We assign an error of 3\% for this. 
The MC statistical uncertainty and modeling with eight $\mpl$ bins 
contributes a 4\% error
in the branching fraction determination.  
The error on the number
of total $B\overline{B}$ pairs is determined to be 1\%. 
The error of the branching fraction for $\Lambda \to p\pi^-$   
is 0.8\%~\cite{PDG}.

\begin{figure}[htb]
\centerline{
\epsfig{file=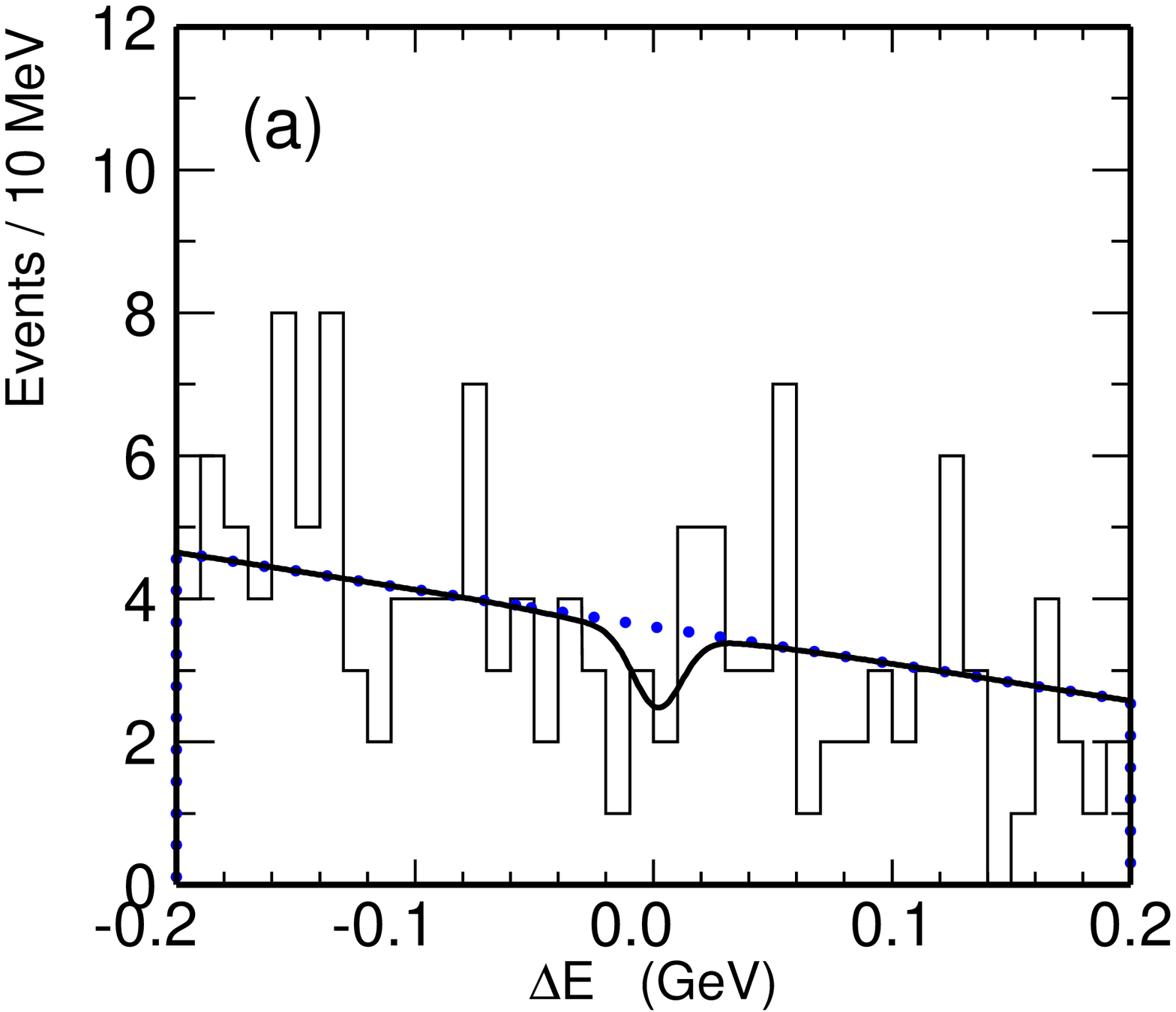,width=2.55in}
\epsfig{file=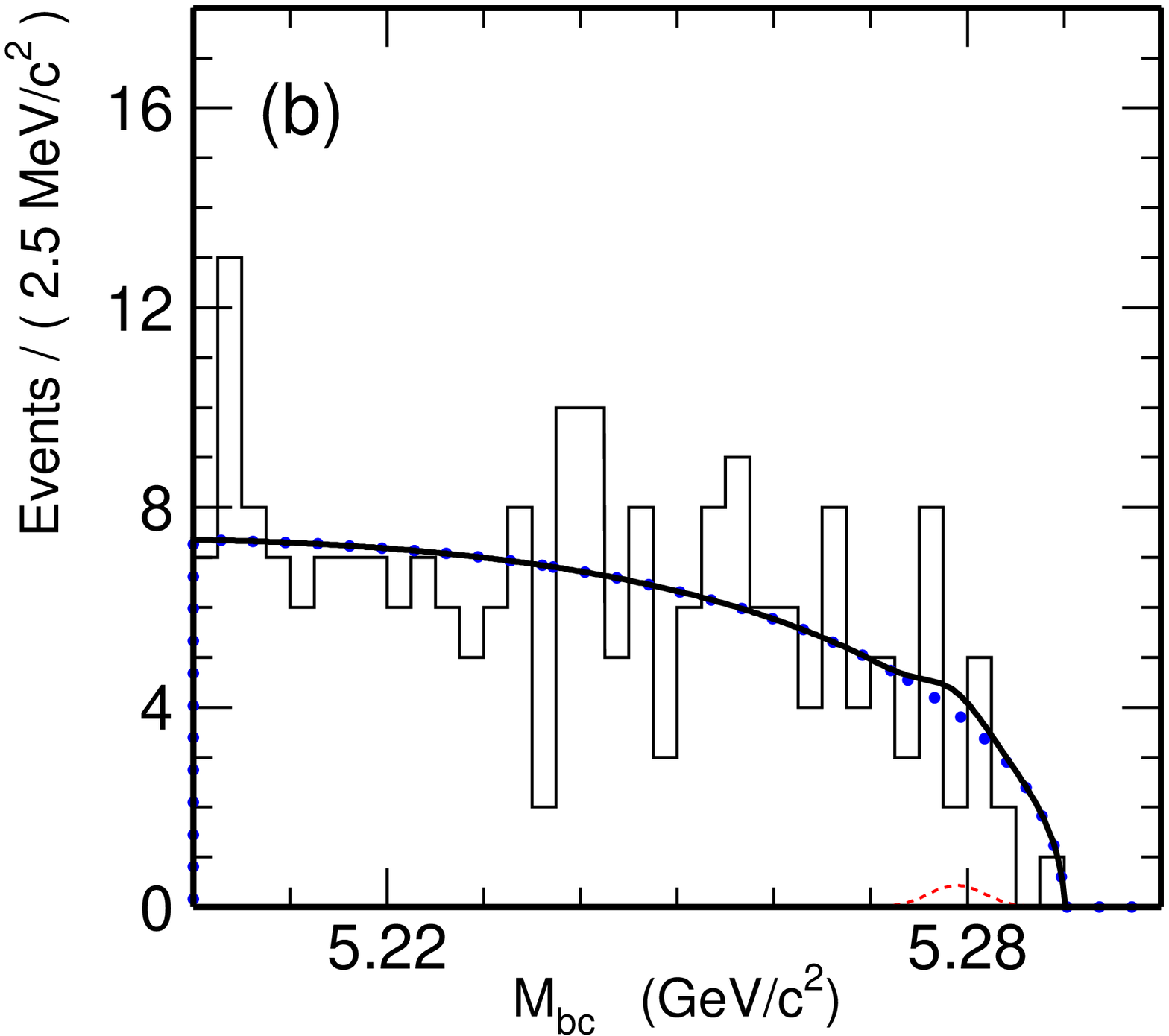,width=2.482in}    
}
\caption{The (a) $\de$ and (b) $\mb$
distributions for $B^0 \to \plk$ candidates.
The solid line represents 
the fit results.}      
\label{fg:plk}
\end{figure}

The tracking systematic error is estimated to be 8\% by summing the correlated
errors of 2\% per charged track. The particle identification error is estimated
to be 8\% by summing the correlated errors of 3\% per proton identification 
and 2\% for the primary pion identification. 
Then we combine them in quadrature along with other uncorrelated
errors to determine a total systematic error of 14\%. 

\begin{figure}[htb]
\centerline{
\epsfig{file=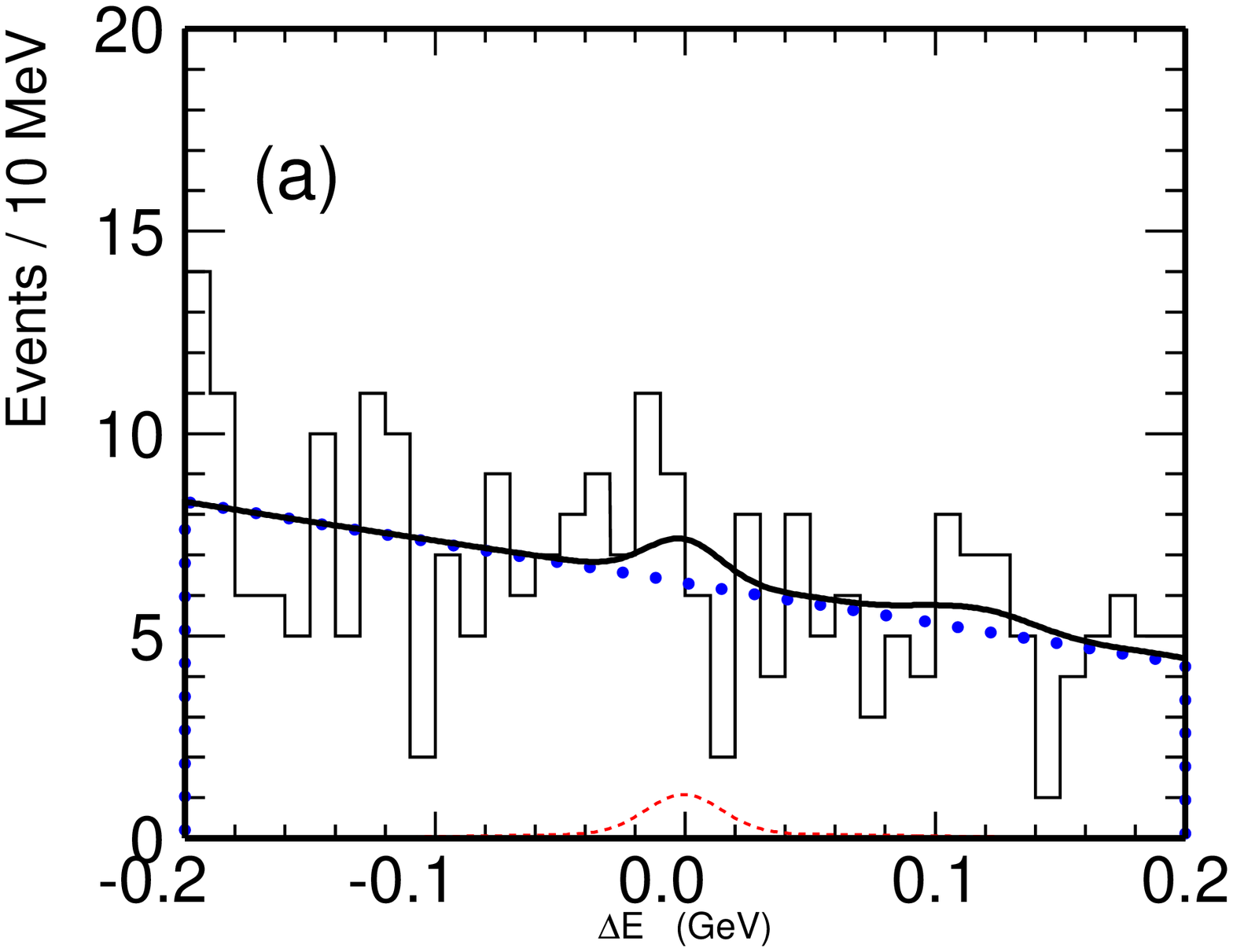,width=2.55in, height=2.873 in}
\epsfig{file=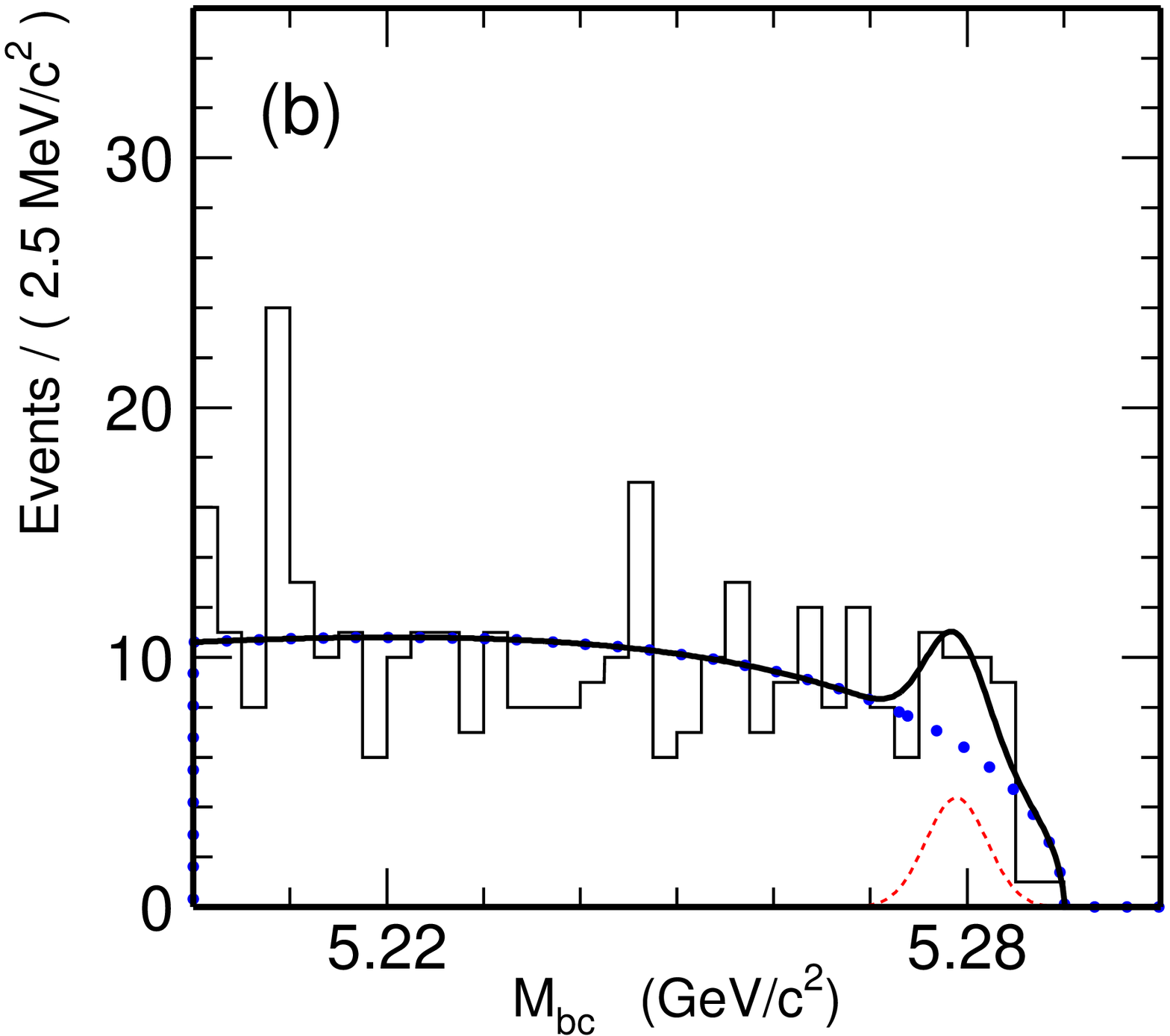,width=2.55in}    
}
\caption{The (a) $\de$ and (b) $\mb$
distributions for $B^0 \to \psigpi$ candidates. Note that events from 
$B^0 \to \plpi$ decay can feed into the high $\de$ region ($\sim$ 0.1 GeV). 
The corresponding  distribution is obtained from the 
$\plpi$ MC and included in the fit.
}   
\label{fg:psigpi}
\end{figure}

We also search for the decay modes $B^0 \to \plk$ and $B^0 \to \psigpi$.
The $\mb$ and $\de$ distributions with fit projections 
are shown in Figs.~{\ref{fg:plk}} and~{\ref{fg:psigpi}}. 
With the signal region extended to  
$|\de|< 0.04$ GeV and $\mb>5.27$ GeV/$c^2$,
no significant signal is found. 
We use the fit results to estimate the expected background, 
and compare this  with the observed number of events  
in the signal region 
in order to set the upper limit on the 
yield at the 90\% confidence level~\cite{Highland,Gary,Conrad}. 
This procedure takes systematic uncertainties into account.
The estimated backgrounds are $28.9 \pm 2.6$ and $50.5 \pm 4.0$, 
the numbers of 
observed events are 26 and 56, the systematic uncertainties are
14\% and 28\%, 
and the upper limit yields are 8.3 and 22.4 
for $\plk$ and $\psigpi$, respectively. 
We estimate the efficiencies from a phase space MC sample. 
The 90\% confidence-level upper limits 
for the branching fractions are  
${\mathcal B}(B^0 \to \plk) < 8.2 \times 10^{-7}$ and
${\mathcal B}(B^0 \to \psigpi) < 3.8 \times 10^{-6}$.

Following our observation of the $B^+\to p\bar pK^+$ mode,
some authors~\cite{CY,CHT} predicted a  much smaller  
branching fraction ($< 10^{-6}$) for the $B^0\to \plpi$ mode,
but a relatively sizable $B^0\to \psigpi$.
Although the predicted rates are not borne out by our present findings,
the threshold peaking behavior shown in Fig.~ \ref{fg:phase}
was anticipated~\cite{HS,rhopn,CHT}.

In summary, we have performed a search for
the rare baryonic decays $B^0 \to \plpi$, $\plk$, and $\psigpi$ 
with $85.0 \pm 0.5$ million $B\bar{B}$ events.
A clear signal is seen in the $\plpi$ mode, 
and we measure a branching fraction of
${\mathcal B}(B^0 \to \plpi) = (3.97^{+1.00}_{-0.80} \pm 0.56) \times 10^{-6}$.
The other two modes are not  seen,
and we set 90\% confidence-level upper limits of 
${\mathcal B}(B^0 \to \plk) < 8.2 \times 10^{-7}$ and
${\mathcal B}(B^0 \to \psigpi) < 3.8 \times 10^{-6}$.

\clearpage

We wish to thank the KEKB accelerator group for the excellent
operation of the KEKB accelerator.
We acknowledge support from the Ministry of Education,
Culture, Sports, Science, and Technology of Japan
and the Japan Society for the Promotion of Science;
the Australian Research Council
and the Australian Department of Industry, Science and Resources;
the National Science Foundation of China under contract No.~10175071;
the Department of Science and Technology of India;
the BK21 program of the Ministry of Education of Korea
and the CHEP SRC program of the Korea Science and Engineering Foundation;
the Polish State Committee for Scientific Research
under contract No.~2P03B 17017;
the Ministry of Science and Technology of the Russian Federation;
the Ministry of Education, Science and Sport of the Republic of Slovenia;
the National Science Council and the Ministry of Education of Taiwan;
and the U.S.\ Department of Energy.


\begin{thebibliography}{99}
\bibitem{conjugate} {Throughout this report, 
inclusion of charge conjugate mode is always implied
unless otherwise stated.}
\end{thebibliography}
\end{document}